\documentclass[pre,twocolumn,amsmath,amssymb,superscriptaddress]{revtex4-1}

\usepackage{graphicx}

\def\tr{\mbox{tr}\,}

\newcommand{\I} {\mbox{Im}\,}

\begin{document}

\title{Characterizing correlations with full counting statistics: \\
classical Ising and quantum XY spin chains}

\author{Dmitri~A.~Ivanov}
\affiliation{Institute for Theoretical Physics, ETH Z\"urich,
8093 Z\"urich, Switzerland}
\affiliation{Institute for Theoretical Physics, University of Z\"urich, 
8057 Z\"urich, Switzerland}

\author{Alexander G.~Abanov}
\affiliation{Department of Physics and Astronomy,
Stony Brook University,  Stony Brook, NY 11794, USA}


\begin{abstract}
We propose to describe correlations in classical and quantum systems
in terms of full counting statistics of a suitably chosen discrete observable.
The method is illustrated with two exactly solvable examples: the classical one-dimensional 
Ising model and the quantum spin-1/2 XY chain. For the one-dimensional Ising model,
our method results in a phase diagram with two phases distinguishable by the long-distance
behavior of the Jordan--Wigner strings. For the anisotropic spin-1/2 XY chain in a
transverse magnetic field, we compute the full counting statistics of the magnetization 
and use it to classify quantum phases of the chain. The method, in this case, 
reproduces the previously known phase diagram. We also discuss the relation
between our approach and the Lee--Yang theory of zeros of the partition function.
\end{abstract}

\maketitle

\section{Introduction}

Thermodynamic phases are traditionally described by
correlations of local observables. In the
simplest case of phase transitions associated with symmetry
breaking, phases are distinguished by the expectation
value of a local order parameter. In more subtle
situations (e.g., Kosterlitz--Thouless phase transition),
it is the decay of correlations at large distances 
that distinguishes between the phases.
In recent years, it was realized that other,
more sophisticated characteristics of correlations may be
useful: e.g., the notion of ``topological order'' 
(involving nonlocal order parameters)
or entanglement entropy (in the case of quantum systems).

In the present work, we consider yet another {\em nonlocal}
characteristics  of correlations based on the 
full-counting-statistics (FCS) approach 
\cite{CD2007} (a related problem of order-parameter statistics was
studied in Ref.~\onlinecite{OP-FCS}). It was pointed out recently 
(in the context of temporal correlations) that
analytical properties of the {\em extensive part} of FCS 
may be used to distinguish
between different thermodynamic phases \cite{IA2010}. 
Related phase-transition effects in FCS were also discussed
in various contexts in Refs.~\onlinecite{Garrahan,Sukhorukov,vonOppen,Lesovik}.
Here we apply the idea of Ref.~\cite{IA2010} 
to spatial correlations and 
illustrate it with two examples: the classical Ising and the quantum XY spin chains
(see also  Ref.~\onlinecite{AIQC2011} for an example of one-dimensional 
free fermions, which do not exhibit any phase transition).

\section{FCS characterization of thermodynamic phases}

In our approach, a thermodynamic phase is characterized by
the singularities of the extensive part of a suitably defined
FCS generating function $\chi_0(\lambda)$. This construction
is applicable to any infinite system (either classical or
quantum, not necessarily one-dimensional) 
which is periodic in space and possesses an extensive
observable taking {\em quantized discrete values}
(e.g., the number of particles or
the projection of total spin on a given axis). Consider
a large subsystem $\Sigma$ containing $N$ unit cells of the
infinite system. Let $Q$ be our discrete observable restricted to
this subsystem and normalized to
take integer values. Then one can construct the FCS
generating function for the observable $Q$,
\begin{equation}
\chi_\Sigma(\lambda) = \langle e^{i\lambda Q} \rangle = \sum_m P_m e^{i\lambda m}\, ,
\label{chi-sigma}
\end{equation}
where the sum is taken over all integer numbers $m$ and $P_m$ 
is the probability for the observable $Q$ to take the
value $m$. The generating function $\chi_\Sigma(\lambda)$ has
the form of a partition function \cite{IA2010} and therefore must
depend exponentially on the size of the system $\Sigma$ \cite{sloppy}:
\begin{equation}
\chi_\Sigma(\lambda) \propto \chi_0(\lambda)^N \, .
\label{chi0-definition}
\end{equation}
Here $\chi_0(\lambda)$ plays the role of 
the {\it extensive part} of $\chi_\Sigma(\lambda)$. It is periodic 
in $\lambda$ (with period $2\pi$), but does not have to be smooth, and 
it is the singularities of $\chi_0(\lambda)$ at real values of $\lambda$
that we propose to use as a characteristics of the thermodynamic
phase \cite{IA2010}.

\begin{figure}
\centerline{\includegraphics[width=0.4\textwidth]{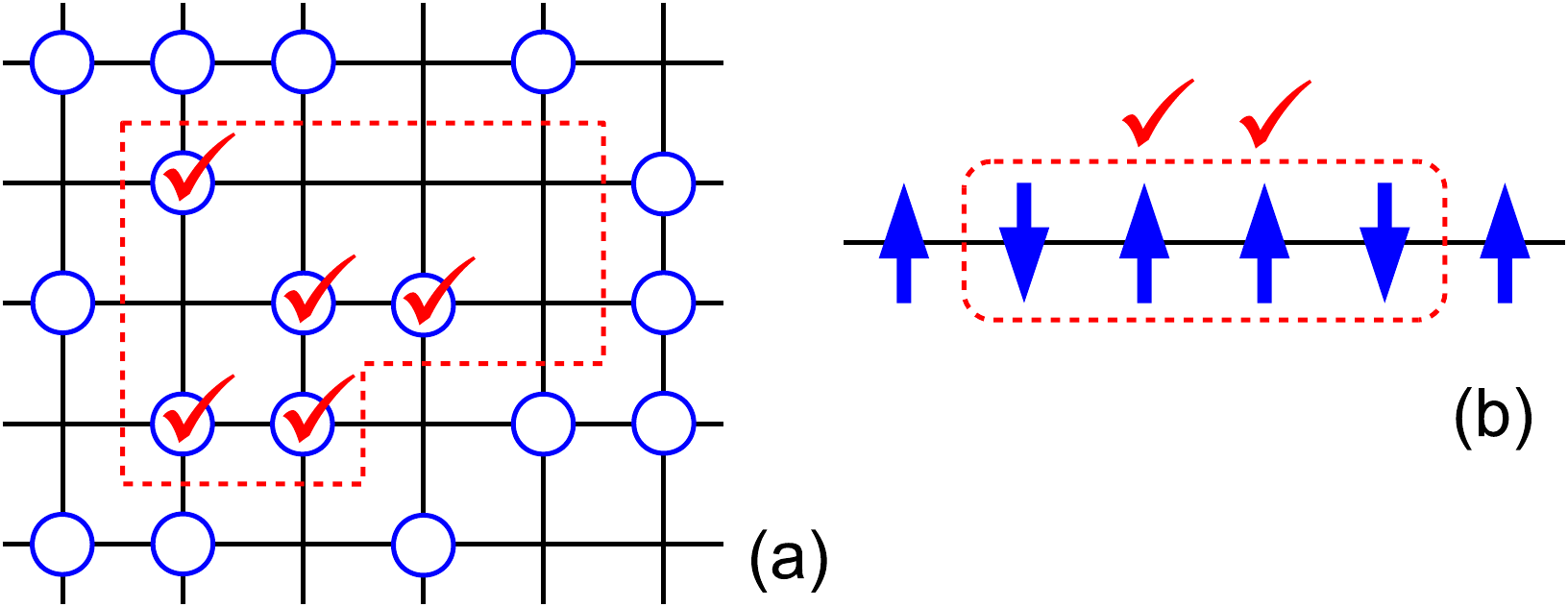}}
\caption{Two examples of FCS in statistical or quantum
systems. In both examples, the subsystem $\Sigma$ is encircled by a dashed line.
{\bf (a)} Particles on a lattice, $Q$ is the number of particles
(in the configuration shown, $Q=5$). {\bf (b)} Spin-1/2 chain,
$Q$ is the number of up spins (in the configuration shown, $Q=2$).}
\label{fig:everything}
\end{figure}

Note that the definition above is quite general and applies to
a vast number of statistical and quantum problems (in many situations,
one even has a choice between different possible observables $Q$).
In Fig.~\ref{fig:everything}, we show two such examples: particles on
a lattice (with $Q$ being the number of particles) and a spin-1/2 chain
(with $Q$ being the number of up spins).

Our FCS construction is related to other characteristics of correlations.
The analytic continuation $\lambda \to i\infty$ produces the 
``emptiness formation probability'' (EFP) \cite{Korepin-book,AF0305}. On the other hand, 
in quantum systems of noninteracting fermions, FCS
is known to be related to the entanglement \cite{entanglement}.

Singularities in $\chi_0(\lambda)$ is a subtle characteristics of
FCS. In Ref.~\onlinecite{IA2010} we argued that they are related to pre-exponential
factors in staggered cumulants of the observable $Q$ 
(if the singularity occurs at $\lambda=\pi$). 
In one-dimensional systems, we can interpret $\ln\chi_0(\lambda)$ as the inverse correlation
length of the Jordan--Wigner string $\exp(i\lambda Q)$ 
(such correlation functions
were also discussed in the context of integrable systems \cite{Korepin-book}). In some 
situations, Jordan--Wigner strings may be related to
physical quantities which are directly observable, e.g.,
spin correlations in the example of the XY chain below.

We propose to classify thermodynamic phases by the number and
type of singularities of $\chi_0(\lambda)$ at real values of $\lambda$.
Thus obtained {\em counting phases} do not have to exactly
reproduce the conventional phase diagram.
Below, we illustrate this proposal 
with two examples.

\begin{figure}
\centerline{\includegraphics[width=0.45\textwidth]{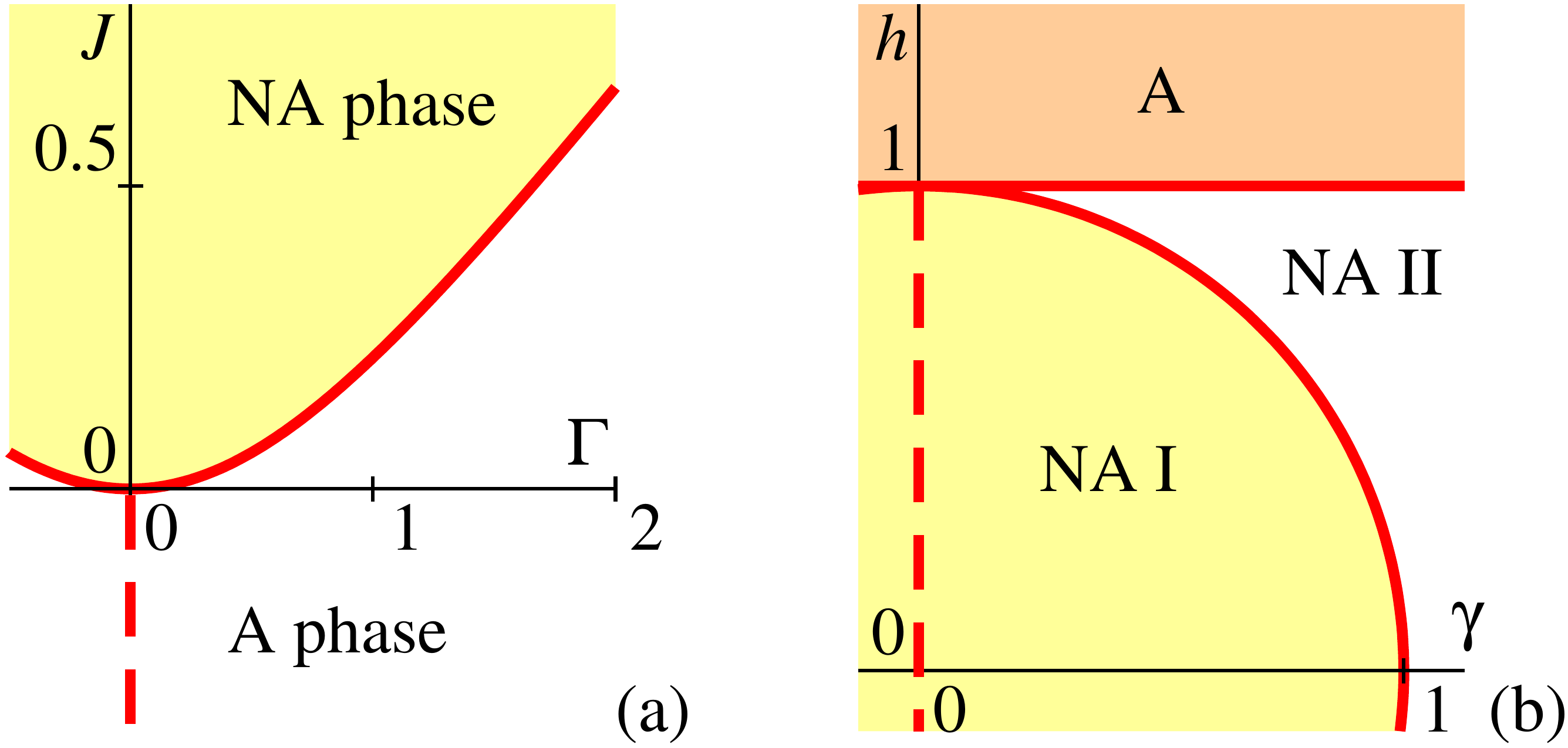}}
\caption{
{\bf (a)} The counting phase diagram of the classical Ising chain.
The shaded upper region is the ``nonanalytic'' (NA) phase, and the
lower region is the ``analytic'' (A) phase \cite{singular-line}.
The phase diagram is symmetric with
respect to the change of sign of $\Gamma$.
{\bf (b)} The counting phase diagram of the spin-1/2 XY
chain at zero temperature.
The three phases are: Nonanalytic I, Nonanalytic II, and Analytic.
On the dashed line $\gamma=0$, $|h|<1$, the dependence of FCS on $\gamma$
is nonanalytic. The phase diagram is symmetric
with respect to the (independent) changes of sign
of $\gamma$ and $h$.
}
\label{fig:two-phase-diagrams}
\end{figure}

\section{One-dimensional Ising model}

Our classification results in a nontrivial phase diagram for the
one-dimensional Ising model. We consider the classical Ising chain
in an external field described by the Hamiltonian
\begin{equation}
H=J\sum_j \sigma_j \sigma_{j+1} + \Gamma \sum_j \sigma_j\, ,
\end{equation}
where the Ising spins $\sigma_j$ take values $\pm 1$ and the
statistical weights of spin configurations are given by $\exp(-H)$
(the temperature is incorporated in the parameters
$J$ and $\Gamma$). This model is equivalent to the ``weather model''
considered in Ref.~\onlinecite{IA2010}, and one finds two different counting phases: 
``analytic'' (A) and ``nonanalytic'' (NA). In the NA phase,
$\chi_0(\lambda)$ has a singularity at $\lambda=\pi$. The phase
diagram is shown in Fig.~\ref{fig:two-phase-diagrams}a, in coordinates
$\Gamma$ and $J$ \cite{weather-Ising,singular-line}. The phase-transition line is given by
\begin{equation}
\cosh\Gamma = e^{2J}\, .
\end{equation}

This counting phase transition has a simple physical interpretation. If one 
considers the Jordan--Wigner string
\begin{equation}
V_\pi(j)=\prod_{k=j_0}^{j} \sigma_k
\label{Jordan-Wigner-Ising}
\end{equation}
(with respect to some reference site $j_0$), then the counting phase transition
corresponds to a
nonanalyticity of the correlation length of the exponentially decaying correlation 
function $\langle V_\pi(0) V_\pi(j) \rangle$, as a function of the parameters 
$J$ and $\Gamma$. In the A phase, the correlation function
$\langle V_\pi(0) V_\pi(j) \rangle$ exhibits a pure exponential (at $\Gamma<0$)
or a staggered-exponential (at $\Gamma>0$) decay as a function of $j$, while 
in the NA phase there are additional {\em incommensurate oscillations} in $j$.
Note that this counting phase transition is not a thermodynamic
phase transition in the usual sense: in fact, thermodynamic phase
transitions are not possible in statistical one-dimensional systems with
local interactions.
The thermodynamic partition function of the Ising model does not have 
a singularity at the counting phase transition, but non-local Jordan-Wigner-type 
correlations (\ref{Jordan-Wigner-Ising}) do. This example illustrates a clear
distinction between counting and conventional thermodynamic
phase transitions.

Finally, we remark that the counting phase diagram of the one-dimensional Ising model
found in this Section can also be understood in terms of the Lee--Yang theory of
zeros of the partition function. Details of this connection are presented in Section
\ref{sec:Lee-Yang} below.

\section{Spin-1/2 XY chain}
\label{sec:spin-chain}

We now turn to a quantum example where a nontrivial counting phase diagram may be explicitly
constructed: the spin-1/2 XY chain in a transverse magnetic field.
The Hamiltonian of the system is \cite{LSM61}
\begin{equation}
{\hat H}=\sum_j \left(\frac{1+\gamma}{2}\, \sigma_j^x \sigma_{j+1}^x +
\frac{1-\gamma}{2}\, \sigma_j^y \sigma_{j+1}^y - h\, \sigma_j^z \right) \, .
\end{equation}
Without loss of generality, we assume $\gamma \ge 0$ and $h \ge 0$.
We are interested in the counting phase diagram with respect to the number of up spins
(as the observable $Q$ in our construction) at zero temperature
(in the particular case of $\gamma=1$, the FCS in this system was
studied in Ref.~\onlinecite{CD2007}). 
By the Jordan--Wigner
transformation, this model can be mapped onto a quadratic fermionic system \cite{LSM61,BM71}, and
the generating function $\chi_\Sigma(\lambda)$ for a subchain of $N$ sites can be
written as a $N\times N$ Toeplitz determinant. By a simple extension of the
derivation in Ref.~\onlinecite{AF0305}, we find
\begin{equation}
\chi_\Sigma(\lambda) = \det_{1\le j \le k \le N} \int_0^{2\pi} \frac{dq}{2\pi}\, 
\sigma(q,\lambda) e^{iq(j-k)}
\label{Toeplitz-det}
\end{equation}
with the {\em symbol} $\sigma(q,\lambda)$ of the Toeplitz determinant given by
[cf.\ Eq.~(17) of the first paper of Ref.~\onlinecite{AF0305}]
\begin{equation}
\sigma(q,\lambda)=\frac{1+e^{i\lambda}}{2} + \left( \frac{1-e^{i\lambda}}{2} \right)
\frac{\cos q - h + i \gamma \sin q}{\sqrt{(\cos q - h)^2 + \gamma^2 \sin^2 q}} \, .
\label{sigma-symbol}
\end{equation}

The exponential asymptotic dependence
of such determinants on $N$ is given by the Szeg\H{o} formula \cite{Szego15}. It immediately
produces the result for $\chi_0(\lambda)$:
\begin{equation}
\chi_0(\lambda)=\exp \int_0^{2\pi} \frac{dq}{2\pi} \, \ln \sigma(q,\lambda)\, ,
\label{Szego-1}
\end{equation}
provided that $\sigma(q,\lambda)$ has zero winding of the complex phase as $q$ 
varies from $0$ to $2\pi$. In the integral (\ref{Szego-1}), the
branch of the logarithm is chosen by the analytic continuation along
the real axis of $q$, and the zero-winding condition implies that 
$\ln \sigma(2\pi,\lambda)= \ln \sigma(0,\lambda)$ under such an analytic
continuation.

For some values of $\gamma$, $h$, and $\lambda$, however, the symbol 
(\ref{sigma-symbol}) may have winding number one [so that 
$\ln \sigma(2\pi,\lambda)= \ln \sigma(0,\lambda) + 2 \pi i$]. In this
case, a modification of the Szeg\H{o} formula applies \cite{FH69}:
\begin{equation}
\chi_0(\lambda)= - \exp \left( \int_0^{2\pi} \frac{dq}{2\pi} \, 
\ln \left[\sigma(q,\lambda) \, e^{-iq} \right] + i q_0 \right) \, ,
\label{Szego-2}
\end{equation}
where $q_0$ is the location of the singularity of $\sigma(q,\lambda)$ in the
upper half plane of $q$ with the smallest imaginary part.

\begin{figure}
\centerline{\includegraphics[width=0.4\textwidth]{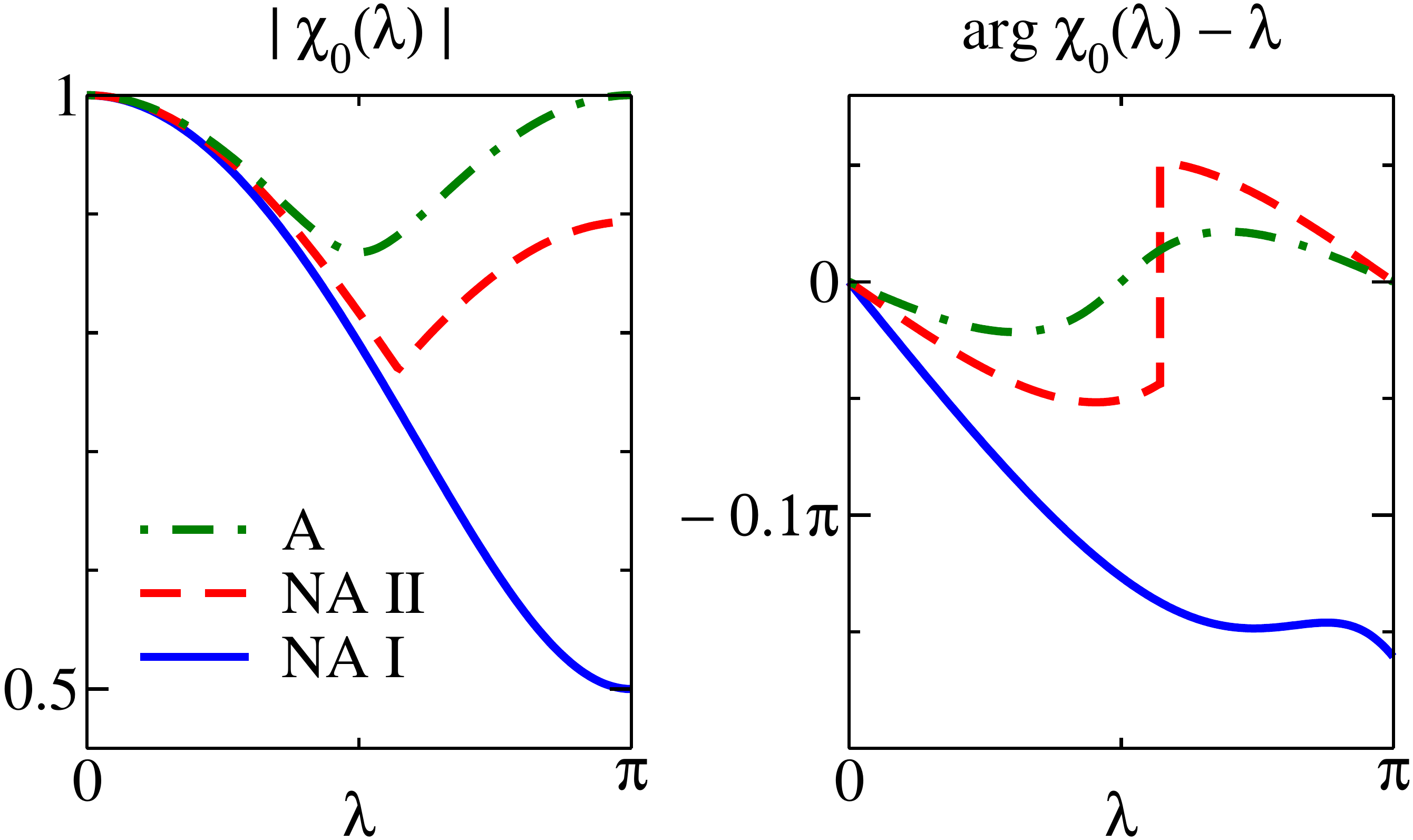}}
\caption{The absolute value and the phase of the generating function
$\chi_0(\lambda)$ at typical points in each of the three counting phases
of the XY chain. For better visualization, the linear part $\lambda$
is subtracted from the phase of $\chi_0(\lambda)$. In each panel,
the three curves are (from bottom to top): 
NA I phase (solid line), $\gamma=0.6$, $h=0.7$;
NA II phase (dashed line), $\gamma=0.5$, $h=0.95$; 
A phase (dash-dotted line), $\gamma=0.5$, $h=1.01$.
}
\label{fig:re-im}
\end{figure}

A tedious, but straightforward application of 
Eqs.~(\ref{Szego-1}) and (\ref{Szego-2}) 
allows us to calculate explicitly $\chi_0(\lambda)$
at all values of $\gamma$ and $h$. As a result, we find three
counting phases shown in Fig.~\ref{fig:two-phase-diagrams}b: two nonanalytic phases and
an analytic one.
Note that the same phase diagram appeared previously
in the analysis of spin correlations \cite{BM71} and of EFP \cite{AF0305}. 
The generating functions  $\chi_0(\lambda)$ at typical points in each of the three phases are
shown in Fig.~\ref{fig:re-im}. Below we summarize some properties of 
these phases in terms of FCS.

{\em Nonanalytic I phase} (NA I): $\gamma^2 + h^2 < 1$. In this phase, we find
(assuming $\lambda \in [-\pi,\pi]$, $\gamma \ge 0$, $h \ge 0$) for the absolute
value of $\chi_0(\lambda)$:
\begin{equation}
| \chi_0(\lambda) | = \sqrt\frac{1+\gamma\cos\lambda}{1+\gamma}\, .
\label{NA1-Re}
\end{equation}
Expressions for the phase of $\chi_0(\lambda)$ following from Eqs.~(\ref{Szego-1})
and (\ref{Szego-2}) are lengthy for all the three phases, and we do not present them here, 
except in several particular cases where
they can be considerably simplified.
In the NA I phase, the only singularity of $\chi_0(\lambda)$ is a phase
jump at $\lambda=\pi$:
\begin{equation}
\I \ln \chi_0(\pm\pi) = \pm \left(\pi - \arccos \frac{h}{\sqrt{1-\gamma^2}} \right)\, .
\end{equation}

In several special cases, $\chi_0(\lambda)$ takes a particularly simple form.
At $\gamma=0$, 
\begin{equation}
\chi_0(\lambda)= \exp \left[ i\lambda \left(1-\frac{1}{\pi}\arccos h \right) \right]
\end{equation}
[in this case, the spin chain is equivalent to free fermions with the density
$1-(1/\pi)\arccos h$].

At $h=0$,
\begin{equation}
\chi_0(\lambda)=e^{i\lambda/2} \sqrt\frac{1+\gamma \cos \lambda}{1+\gamma}\, .
\end{equation}

At the phase boundary $\gamma^2 + h^2 =1$,
\begin{equation}
\chi_0(\lambda)= p\, e^{i\lambda} + 1-p\, , \qquad
p=\frac{1}{2}\left(1 + \sqrt\frac{1-\gamma}{1+\gamma} \right)\, ,
\label{NA12-boundary}
\end{equation}
which corresponds to independent spins with the probability $p$
of pointing up \cite{BM71}.

Note that Eq.~(\ref{NA1-Re}) is not even in $\gamma$, and therefore $\chi_0(\lambda)$
depends nonanalytically on $\gamma$ across the line $\gamma=0$.

{\em Nonanalytic II phase} (NA II): $\gamma^2 + h^2 > 1$, $|h|<1$. In this phase
(again assuming $\lambda \in [-\pi,\pi]$, $\gamma \ge 0$, $h \ge 0$), $\chi_0(\lambda)$
has phase jumps at points $\pm \lambda_c$ given by
\begin{equation}
\cos\lambda_c = - \frac{1-z_1^2}{\gamma (1+z_1^2)}\, ,
\qquad z_1=\frac{h+\sqrt{\gamma^2 + h^2 -1}}{1+\gamma}\, .
\end{equation}
For the absolute value of $\chi_0(\lambda)$, we find
\begin{equation}
|\chi_0(\lambda) | = 
\begin{cases}
\sqrt\frac{1+\gamma\cos\lambda}{1+\gamma}\, , &
\text{$|\lambda|<\lambda_c$} \, , \\
z_1 \, \sqrt\frac{1-\gamma\cos\lambda}{1+\gamma}\, , &
\text{$|\lambda|>\lambda_c$} \, .
\end{cases}
\label{NA2-Re}
\end{equation}
Once again, we do not present here full expressions for the phase of $\chi_0(\lambda)$.
The phase jump at $\lambda_c$ is given by
\begin{equation}
\I \ln \frac{\chi_0(\lambda_c+0)}{\chi_0(\lambda_c-0)} = \arccos \frac{h(1+z_1^2)}{2 z_1}
\end{equation}
(this jump tends to zero at the phase boundary). At the boundary with the NA I phase, 
$\lambda_c \to \pi$, and $\chi_0(\lambda)$ is given by Eq.~(\ref{NA12-boundary}).
At the boundary with the Analytic phase $h=1$, one finds $\lambda_c \to \pi/2$.
In the whole NA II phase, $\I \ln \chi_0 (\pm \pi)= \pm \pi$, which implies that
$\chi_0(\lambda)$ is smooth at $\lambda=\pi$.

{\em Analytic phase} (A): $|h|>1$. In this phase, $\chi_0(\lambda)$ has
no singularities in $\lambda$. For its absolute value we find:
\begin{equation}
| \chi_0(\lambda) | = 
\sqrt\frac{h + \sqrt{\gamma^2 \cos^2 \lambda + h^2 -1 }}{h + 
\sqrt{\gamma^2 + h^2 -1 }}\, .
\end{equation}
Throughout this phase, $\I \ln \chi_0 (\pm \pi)= \pm \pi$ (just like
in the NA II phase). At $\gamma=0$, the generating function takes
the particularly simple form
$\chi_0(\lambda)=i\lambda$ [a completely
filled band of free fermions].

We notice a remarkable property of the A phase: throughout this phase,
$\chi_0(\pi)=-1$. This means that the correlations of the Jordan--Wigner
operators (\ref{Jordan-Wigner-Ising}) [where $\sigma_k$ denote now
the $z$ components $\sigma_k^z$] decay slower than exponentially.
This property was described in Ref.~\onlinecite{CD2007}
as ``confinement of dual domain walls'', and it can also be understood 
in the fermionic language.
At $|h|>1$, fermions form a completely filled (or completely empty) band, 
and the anisotropy terms (involving $\gamma$) introduce local Cooper
pairs, which, however, change the number of particles by two. Thus
the expectation value of the Jordan--Wigner parity string 
$\langle V_\pi(0) V_\pi(j) \rangle$ is only affected by Cooper
pairs intersecting one of the end points of the string ($0$ or $j$).
But, since pairs are local, the number of such pairs remains
finite for long strings, which leads to a saturation 
of the correlations $\langle V_\pi(0) V_\pi(j) \rangle$
at large $j$.

The counting phase diagram of the XY chain (Fig.~\ref{fig:two-phase-diagrams}b)
coincides with that obtained from spin correlations:
different phases are distinguished by the transverse long-range order, 
presence or absence of incommensurate oscillations in spin correlations 
and by pre-exponential factors in exponentially decaying spin correlations \cite{BM71}.
This coincidence is not surprising, since,
by the Jordan--Wigner transformation, transverse spin operators $\sigma_j^{\pm}$
are represented by the product of the string operator (\ref{Jordan-Wigner-Ising})
and a fermion operator \cite{LSM61}. As a consequence, transverse spin correlations
are given by the Toeplitz determinants which differ from Eqs.\ (\ref{Toeplitz-det})
and (\ref{sigma-symbol}) only by a shift of the winding number (and by
fixing $\lambda=\pi$) \cite{BM71}. 

The phase diagram in Fig.~\ref{fig:two-phase-diagrams}b also 
resembles those based on EFP in Ref.~\onlinecite{AF0305}
and on entanglement entropy in Ref.~\onlinecite{XY-entanglement}.
Indeed, 
the EFP is given by the same Toeplitz determinant (\ref{Toeplitz-det}), 
(\ref{sigma-symbol}) with $\lambda \to i \infty$ \cite{AF0305}, and
the entanglement entropy depends on the spectrum of a closely
related block Toeplitz matrix \cite{XY-entanglement}.
Therefore the phase boundaries
which are determined by the geometry of the square-root branching points in 
Eq.~(\ref{sigma-symbol}) coincide in all the three
problems. However the FCS classification contains additional details
related to the positions of $\lambda$-dependent logarithmic branching 
points in Eq.~(\ref{Szego-1}).

Finally, we remark that, even though the XY spin chain considered in our
work maps onto a quadratic fermionic system, it does not obey the theorem
on factorization of FCS for noninteracting fermions of Ref.~\onlinecite{AI0809}:
the phase NA II with a singularity at an intermediate value of $\lambda$
would not be allowed by that theorem. 
The reason for this discrepancy is that the corresponding fermionic system
contains pairing terms \cite{LSM61}, and the theorem of Ref.~\onlinecite{AI0809}
is not, in general, valid for quadratic Hamiltonians with pairing
(see also discussion in the supplementary material of the last paper of 
Ref.~\onlinecite{entanglement}). Thus the NA II phase presents
an example of an {\it interacting} system with singularitites of 
$\chi_0(\lambda)$ shifted away from $\lambda=\pi$.

\section{Relation to Lee--Yang zeros}
\label{sec:Lee-Yang}

Our theory of counting phase transitions is closely related to the approach
of Lee and Yang considering zeros of the partition function in the complex
plane of a parameter of the model (fugacity or magnetic field) \cite{Lee-Yang}. 
Namely, in many (but not all) situations, the counting phase transitions are determined
by the locations of zeros of the partition function in the complex fugacity
(or magnetic-field) plane.

Consider first the case of a classical system. Then the generating function
(\ref{chi-sigma}) can be understood as the partition function 
of the system with the imaginary part
$-i\lambda$ added to the chemical potential dual to the observable $Q$ in
the subsystem $\Sigma$. Thus we study the same analytic continuation of
the partition function as Lee--Yang's one, 
but with a different order of thermodynamic limits. In Lee--Yang theory, 
the chemical potential acquires a uniform imaginary component in the whole 
system, and then the system size tends to infinity. In contrast, in our 
construction, the system size is infinite from the very beginning,
it is only the subsystem $\Sigma$ where the chemical potential has
an imaginary part, and the thermodynamic limit is defined by expanding
the subsystem $\Sigma$ within the same infinite system.

We believe that in most situations this difference in the order 
of the thermodynamic limits is unimportant
and the extensive part of the generating function $\chi_0(\lambda)$
simply equals the Lee--Yang partition function per unit cell of the
system. In this case, the counting phase diagram and the positions
of singularities of $\chi_0(\lambda)$ can be easily read
off the locus of the Lee--Yang zeros in the thermodynamic limit.
The irrelevance of the order of the thermodynamic limits can
be most easily understood in the case of one-dimensional classical systems:
there, both $\chi_0(\lambda)$ and the asymptotic behavior of the
Lee--Yang partition function in the thermodynamic limit are given
by the leading eigenvalue of the corresponding transfer matrix.
This connection to Lee--Yang zeros was also discussed in Ref.~\cite{flindt} 
in the context of ``dynamical phase transitions''.

In the case of a quantum system, the relation to the locus of Lee--Yang zeros 
is more complicated. Here one should distinguish two
possibilities: either the observable $Q$ (defined for the full system) 
commutes with the Hamiltonian $H$ or it does not.

In the case of $Q$ commuting with $H$, the Lee-Yang partition function
$Z_{\rm LY}$ may be related to the FCS generating function for the full system in
the grand-canonical ensemble, just like in the classical case discussed
above:
\begin{equation}
Z_{\rm LY}=\tr e^{-\beta H + i \lambda Q} = \tr e^{-\beta H} e^{i\lambda Q}
= \langle e^{i\lambda Q} \rangle \, \tr e^{-\beta H}
\label{LeeYang-FCS}
\end{equation}
(here $\beta$ is the inverse temperature). Therefore, one may draw the same
relation between the locus of Lee--Yang zeros and counting phase
transitions as in classical systems, provided the order of the thermodynamic 
limits is unimportant. The latter is, however, not always the case, at
least for systems with sufficiently slowly decaying correlations. An example
where the Lee--Yang approach and FCS give different results is one-dimensional
fermions at zero temperature (equivalent to the spin-1/2 XY chain considered
in Section \ref{sec:spin-chain} at $\gamma=0$). There, the FCS approach
leads to a well defined generating function $\chi_0(\lambda)$ \cite{AIQC2011}, 
while the Lee--Yang approach provides neither a good locus of zeros nor a good
thermodynamic partition function per unit cell of the lattice. The question
about the general conditions under which the two orders of the thermodynamic
limit are equivalent, goes beyond the scope of the present paper, and we
leave it for future studies.

Finally, in the case of $Q$ not commuting with $H$ (which applies to
our example of spin-1/2 XY chain in Section \ref{sec:spin-chain} at 
$\gamma\ne 0$), the relation (\ref{LeeYang-FCS}) between the Lee--Yang
partition function and full counting statistics no longer holds, and we
conclude that there is no obvious relation between the two approaches.

\section{Conclusion}

We have proposed a classification scheme of thermodynamic phases and
correlations in general in terms of the analytical properties of the extensive part 
of FCS for a suitably chosen discrete observable. 
Using FCS for describing correlations involves nonlocal observables (\ref{chi-sigma})
and therefore allows us to capture subtle details of correlations inaccessible with
local observables. In this way, {\it counting phases} may be distinguished: sometimes they
coincide with conventional thermodynamic phases, but sometimes one thermodynamic phase
may be further subdivided into several counting phases reflecting differences in statistics
of fluctuations of a certain discrete variable. The physical meaning of counting phase
transitions is often subtle. In some situations it is related to the structure of
zeros of the partition function in Lee--Yang theory, and in one-dimensional situations
we relate FCS physics to the correlations of nonlocal string observables of Jordan--Wigner 
type. Deeper physical implications of these counting phase transitions are still to be
understood.

In our work, we have illustrated our proposal with two simplest examples:
one-dimensional Ising model as a classical example and spin-1/2 XY model as
a quantum one. We have chosen those examples, since they allow an analytical
calculation of the counting phases and an easy comparison with other available
results. Studying counting phases in systems with more complicated interactions 
is interesting but requires more involved analytical or numerical methods.

In general, the role of interactions in counting phase transitions is not yet fully
understood. While it has been shown in Ref.~\onlinecite{AI0809} that
fermionic systems in the absence of interactions can only exhibit 
singularities in $\chi_0(\lambda)$ at real negative $e^{i\lambda}$,
the absence of interactions is not a necessary condition for this property.
Indeed, for the spin-1/2 XY chain at a finite anisotropy $\gamma$ 
(see Section \ref{sec:spin-chain}) both phases with singularities at 
$\lambda=\pi$ and those with singularities at $\lambda\ne\pi$ are present.

Finally, we remark that here we have only focused on analytic properties 
of the generating function $\chi_0(\lambda)$ at real values of $\lambda$. 
It may also be instructive to analyze, more generally, 
the structure of singularities of $\chi_0(\lambda)$ 
in the complex plane of $e^{i\lambda}$. This
would, on one hand, provide a closer connection
to the theory of Lee--Yang zeros \cite{Lee-Yang}, 
and on the other hand, relate it
to the theory of the emptiness formation 
probability \cite{Korepin-book,AF0305}.

\section{Acknowledgments}
The work of A.~G.~A.\ was supported by the NSF under Grant No.\ DMR-1206790.




\begin{thebibliography}{99}

\bibitem{CD2007}
R.~W.~Cherng and E.~Demler, New J.\ Phys.\ {\bf 9}, 7 (2007).
 \\ {\it Quantum noise analysis of spin systems realized with cold atoms.}

\bibitem{OP-FCS}
V.~Eisler, Z.~R\'acz, and F.~van~Wijland, Phys.\ Rev.\ E {\bf 67}, 056129 (2003);
 \\ {\it Magnetization distribution in the transverse Ising chain with energy flux.} \\
A.~Lamacraft and P.~Fendley, Phys.\ Rev.\ Lett.\ {\bf 100}, 165706 (2008).
 \\ {\it Order parameter statistics in the critical quantum Ising chain.}

\bibitem{IA2010}
D.~A.~Ivanov and A.~G.~Abanov, Europhys.\ Lett.\ {\bf 92}, 47008 (2010).
 \\ {\it Phase transitions in full counting statistics for periodic pumping.}

\bibitem{Garrahan}
J.~P.~Garrahan et al, Phys.\ Rev.\ Lett.\ {\bf 98}, 195702 (2007);
 \\ {\it Dynamical first-order phase transition in kinetically constrained models
 of glasses.} \\
J.~P.~Garrahan and I.~Lesanovsky, Phys.\ Rev.\ Lett.\ {\bf 104}, 160601 (2010).
 \\ {\it Thermodynamics of quantum jump trajectories.}

\bibitem{Sukhorukov}
I.~P.~Levkivskyi and E.~V.~Sukhorukov, Phys.\ Rev.\ Lett.\ {\bf 103}, 036801 (2009).
 \\ {\it Noise-induced phase transition in the electronic Mach--Zehnder interferometer.}

\bibitem{vonOppen}
T.~Karzig and F.~von~Oppen, Phys.\ Rev.\ B {\bf 81}, 045317 (2010).
 \\ {\it Signatures of critical full counting statistics in a quantum-dot chain.}

\bibitem{Lesovik}
G.~B.~Lesovik and I.~A.~Sadovskyy, Usp.\ Fiz.\ Nauk {\bf 181}, 1041 (2011)
[Phys.\ Usp.\ {\bf 54}, 1007 (2011)].
 \\ {\it Scattering matrix approach to the description of quantum electrom transport.}

\bibitem{AIQC2011}
A.~G.~Abanov, D.~A.~Ivanov, and Y.~Qian, J.~Phys.~A: Math.\ Theor.\ {\bf 44}, 485001 (2011);
 \\ {\it Quantum fluctuations of one-dimensional free fermions and Fisher--Hartwig formula for 
 Toeplitz determinants.} \\
D.~A.~Ivanov, A.~G.~Abanov, and V.~V.~Cheianov, J.~Phys.~A: Math.\ Theor.\ {\bf 46}, 085003 (2013).
 \\ {\it Counting free fermions on a line: a Fisher--Hartwig asymptotic expansion for
 the Toeplitz determinant in the double-scaling limit.}

\bibitem{sloppy}
In Eq.~(\ref{chi0-definition}), we have include only the main exponential dependence.
The asymptotic behavior of $\chi_\Sigma(\lambda)$ on $N$ may also contain algebraic
pre-exponential terms. 

\bibitem{Korepin-book}
V.~E.~Korepin, N.~M.~Bogoliubov, and A.~G.~Izergin, 
{\it Quantum Inverse Scattering Method and Correlation Functions}, 
Cambridge University Press, Cambridge, UK (1993).

\bibitem{AF0305}
A.~G.~Abanov and F.~Franchini, 	Phys.\ Lett.\ A {\bf 316}, 342 (2003);
 \\ {\it Emptiness formation probability for the anisotropic XY spin chain in a magnetic field.} \\
F.~Franchini and A.~G.~Abanov, J.\ Phys.\ A: Math.\ Gen.\ {\bf 38} 5069 (2005).
 \\ {\it Asymptotics of Toeplitz determinants and the emptiness formation probability for the XY spin chain.}

\bibitem{entanglement}
I.~Klich and L.~Levitov, Phys.\ Rev.\ Lett.\ {\bf 102}, 100502 (2009);
 \\ {\it Quantum noise as an entanglement meter.} \\
H.~F.~Song et al, Phys.\ Rev.\ B {\bf 83}, 161408 (2011);
 \\ {\it Entanglement from charge statistics: exact relations for many-body systems.} \\
H.~F.~Song et al, Phys.\ Rev.\ B {\bf 85}, 035409 (2012).
 \\ {\it Bipartite fluctuations as a probe of many-body entanglement.}

\bibitem{weather-Ising}
The parameters of the weather model $q_r$ and $q_s$ in Ref.~\onlinecite{IA2010} are related to the
Ising parameters $J$ and $\Gamma$ in the present work by
$e^{2\Gamma}=q_r/q_s$ and $e^{4J}=(1-q_r)(1-q_s)/(q_r q_s)$.

\bibitem{singular-line}
The phase diagram in Fig.~\ref{fig:two-phase-diagrams}a shows an additional singular line
$\Gamma=0$, $J<0$, which was omitted in Ref.~\onlinecite{IA2010}. This line
separates two regions of the same analytic phase (one with exponential and
the other with staggered-exponential correlations $\langle V_\pi(0) V_\pi(i) \rangle$).
On this line, the function $\chi_0(\lambda)$ is nonanalytic with a singularity at an 
intermediate value of $\lambda$ (between $0$ and $\pi$).

\bibitem{Lee-Yang}
C.~N.~Yang and T.~D.~Lee, Phys.\ Rev.\ {\bf 87}, 404 (1952);
 \\ {\it Statistical theory of equations of state and phase transitions. 
 I. Theory of condensation.}
T.~D.~Lee and C.~N.~Yang, {\it ibid.} {\bf 87}, 410 (1952).
 \\ {\it Statistical theory of equations of state and phase transitions. 
 II. Lattice gas and Ising model.}

\bibitem{LSM61}
E.~Lieb, T.~Schultz, and D.~Mattis, Ann.\ Phys.\ {\bf 16}, 407 (1961).
 \\ {\it Two soluble models of an antiferromagnetic chain.}

\bibitem{BM71}
E.~Barouch and B.~M.~McCoy, Phys.\ Rev.\ A {\bf 3}, 786 (1971).
 \\  {\it Statistical mechanics of the XY model II: spin-correlation functions.}

\bibitem{Szego15}
G.~Szeg\H{o}, Math.\ Ann.\ {\bf 76}, 490 (1915).
 \\ {\it Ein Grenzwertsatz \"uber die Toeplitzschen Determinanten einer reellen positiven Funktion.}

\bibitem{FH69}
M.~E.~Fisher and R.~E.~Hartwig, Adv.\ Chem.\ Phys.\ {\bf 15}, 333 (1969);
 \\ {\it Toeplitz determinants: some applications, theorems, and conjectures.} \\
R.~E.~Hartwig and M.~E.~Fisher, Arch.\ Ration.\ Mech.\ Anal.\ {\bf 32}, 190 (1969).
 \\ {\it Asymptotic behavior of Toeplitz matrices and determinants.}

\bibitem{XY-entanglement}
A.~R.~Its, B.-Q.~Jin, and V.~E.~Korepin,
J.\ Phys.\ A: Math.\ Gen.\ {\bf 38}, 2975 (2005); 
 \\ {\it Entanglement in XY Spin Chain.} \\
F.~Franchini, A.~R.~Its, B.-Q.~Jin, and V.~E.~Korepin,
J.\ Phys.\ A: Math.\ Theor.\ {\bf 40 }, 8467 (2007).
 \\ {\it Ellipses of constant entropy in the XY spin chain.}

\bibitem{AI0809}
A.~G.~Abanov and D.~A.~Ivanov, Phys.\ Rev.\ Lett.\ {\bf 100}, 086602 (2008);
 \\ {\it Allowed charge transfers between coherent conductors driven by a 
 time-dependent scatterer.} \\
A.~G.~Abanov and D.~A.~Ivanov, Phys.\ Rev.\ B {\bf 79}, 205315 (2009).
 \\ {\it Factorization of quantum charge transport for noninteracting fermions.}

\bibitem{flindt}
C.~Flindt and J.~P.~Garrahan, Phys.\ Rev.\ Lett.\ {\bf 110}, 050601 (2013).
 \\ {\it Trajectory phase transitions, Lee--Yang zeros, and
 high-order cumulants in full counting statistics.}

\end{thebibliography}
\end{document}